\documentclass[a4paper]{jpconf}
\usepackage{graphicx}
\usepackage{cite}
\begin{document}
\title{Relativistic Static Thin Disks of Polarized Matter}

\author{ Anamaria Navarro and F. D. Lora-Clavijo and Guillermo A. Gonz\'alez}
\address{Grupo de Investigaci\'on en Relatividad y Gravitaci\'on,
Escuela de F\'isica, Universidad Industrial de Santander, A. A. 678, Bucaramanga 680002, Colombia.}

\ead{AN: ana.navarro1@correo.uis.edu.co ; FDLC: fadulora@uis.edu.co; GAG: guilllermo.gonzalez@saber.uis.edu.co}


\begin{abstract}
An infinite family of exact solutions of the electrovacuum Einstein-Maxwell equations is presented. The family is static, axially symmetric and describe thin disks composed by electrically polarized material in a conformastatic spacetime. The form of the conformastatic metric allows us to write down the metric functions and the electromagnetic potentials in terms of a solution of the Laplace equation. We find a general expression for the surface energy density of the disk, the pressure, the polarization vector, the electromagnetic fields and the velocity rotation for circular orbits. As an example, we present the first model of the family and show the behavior of the different physical variables.
\end{abstract}


\section{Introduction}

Exact explicit solutions of Einstein equations with axial symmetry have played a crucial role in astrophysical applications of
general relativity. In particular, disk-like configurations have been used to model different kind of configurations like galactic disks, 
accretion disks around compact objects and also to explain the mechanism for extracting energy from the black 
hole to power the launching of the jet. 

The first models of relativistic thin disks were obtained in 1968 by Bonnor and Sackfield   \cite{Bonnor_Sackfield_CommMathPhys8-338} and in by 1969 Morgan and Morgan  \cite{Morgan-Morgan1969}. Since then, many authors have obtained different types of exact solutions of static thin disks 
 \cite{Morgan-Morgan1970, Voorhees_RelativisticDisksI,  Lynden_Pineault_MonNot185-679_I_Relativistic_disk, Letelier-OliveiraJMathPhys28-165-1987, Lemos_SelfSimilarRelativisticDiskswithPressure, BicakLyndenKatzPhysRevD47,Gonzalez:2015qzt}, and stationary thin disks 
 \cite{Lynden_Pineault_MonNot185-695_II_Relativistic_disk, BicakLedvinka-RelativisticDiskKerrMetric}, whereas the stability of this systems was studied in \cite{UjevicLetelierPhysRevD70_2004} using a first order perturbation on the energy-momentum tensor.   
On the other hand, due to the high relevance of electromagnetic fields in general relativity, there have been obtained many solutions that correspond to disks with electric and magnetic fields generated by electrically charged matter and electric currents, see for instance 
 \cite{GarciaReyesGonzalez, LoraOspinaPedraza, GarciaGonzalez2004PhysD96, LetelierExactRelatDisksMagneticFields, VogtLetelierQuantum2004, KatzBicakLynden_Disksourcesforconformastationarymetrics, GonzalezGutierrezOspina, VogtLetelierPhysRevD70_2004, GarciaReyesGonzalez2004ClassQuantum21, GarciaGonzalezTaub}.
 
 Electric or magnetic dipoles have not  been studied very much in the context of exact solutions. The only related articles are  \cite{monopolos,dipolosenschwarchild}, where they analyze the electromagnetic fields generated by electric and magnetic dipole layers in a Schwarzschild spacetime. 
Since there have not been obtained exact solutions of the Einstein-Maxwell that may be interpreted as electric polarized source, in this paper we construct an infinite family of relativistic static thin disks composed by polarized material. In order to do so, we solve the Einstein-Maxwell equations in a conformastatic spacetime using the distributional approach of tensors. Finally, we obtain the general expressions for the surface energy-momentum tensor, the electric field, the polarization vector and analyze them for the first model of the family. 


\section{Einstein Maxwell Field Equations}

The Einstein-Maxwell system of equations for a continuum media, in Heaviside-Lorentz geometrized units \footnote{$c= 8\pi G = \mu_0 = \epsilon = 1$}, can be written as 
\begin{eqnarray}
 G_{\alpha \beta} =  T^M_{\alpha \beta} + T^F_{\alpha \beta} +  T^{FM}_{\alpha \beta} \, ,  ~~~~~ 
 {F^{\alpha \beta}}_{;\beta} = {M^{\alpha \beta}}_{;\beta} \, , \label{eq_EinMax}
\end{eqnarray}
where the components of the electromagnetic tensor $F_{\alpha \beta}$ and the components of the polarization tensor $M_{\alpha \beta}$ are given in terms of the electromagnetic potential $A_{\alpha}$ and the electric polarization vector $P_{\alpha}$ as
\begin{eqnarray}
 F_{\alpha \beta}=A_{\beta, \alpha} - A_{\alpha, \beta}, ~~~~~ M^{\alpha \beta} = P^{\alpha} u^{\beta} - P{^\beta} u^{\alpha}.
\end{eqnarray} 
Here $u_{\alpha}$ corresponds to the four velocity vector of the fluid.  As we can see from equations (\ref{eq_EinMax}), the energy momentum tensor is splitted into three terms \cite{Grot_JMathPhys-11-109_1970}
\begin{eqnarray}
T_{\alpha \beta} &=& T^ {M}_{\alpha \beta} + T_{\alpha \beta}^F + T^{FM}_{\alpha \beta}, \\
T_{\alpha \beta}^F &=& F_{\alpha \mu} {F_{\beta}}^{\mu} - \frac{1}{4}g_{\alpha \beta} F_{\mu \nu} F^{\mu \nu} \, , \label{TF} \\
T^{FM}_{\alpha \beta} &=& -F_{\alpha \mu} {M_{\beta}}^{\mu} \, , \label{TFM}
\end{eqnarray}
which correspond to the matter $T^ {M}_{\alpha \beta}$, the electromagnetic fields $T_{\alpha \beta}^{F}$ and  the electromagnetic interaction with the polarized matter $T_{\alpha \beta}^{FM}$, respectively. It is worth mentioning that the matter energy momentum tensor, the electromagnetic four potential and the electric polarization vector are calculated by using the displace, cut and reflection inverse method \cite{kuzmin1956model}.


\section{Axially Symmetric Solutions for Electrically Polarized Disks}
\label{sec:ASSEPD}

In order to write down the Einstein-Maxwell field equations for electrically polarized infinitesimal thin disks, the line element  for a conformastatic axially symmetric space-time in cylindrical coordinates $ x^a = (t,r,\phi,z)$, can be written as \cite{Synge}
\begin{equation}
 ds^2 = -e^{2\psi} dt^2 + e^{-2\psi} \left( dr^2+r^2d\phi^2 + dz^2 \right) \, , \label{ds2}
\end{equation}
where the metric function $\psi $ depends on the coordinates $(r,z)$. The fact that the solution is not magnetized, the  electromagnetic four potential takes the form $A_\alpha = \left(-\chi,0,0,0\right)$, where the new function $\chi$ also depends on the coordinates $(r,z)$. Based on these, the non-zero component of the polarization vector is given by 
\begin{equation}
P_r =  e^{-\psi} \Pi_{r t} , 
\end{equation}
where $\Pi_{\alpha \beta}$ are the components of the polarization tensor on the disk, which are related with the components of  $M^{\alpha \beta}$ through the relation $M^{\alpha \beta}=\Pi^{\alpha \beta}\delta(z)$, where $\delta(z)$ is the Dirac distribution with support in the surface $z=0$. As we can see, the antisymmetry of the polarization tensor implies that $\Pi_{r t} = -\Pi_{t r} = P_r e^\psi$. 

Solving the Einstein-Maxwell equations (\ref{eq_EinMax}), outside and inside of the disk, we obtain that its surface energy density $\sigma$, pressure $(p_r,p_{\phi},p_z)$ and electric components $(E_{(r)},E_{(z)})$ are
\begin{eqnarray}
\nonumber &&\sigma =e^{\psi} \left( 4\psi_{,z}+\sqrt{2} \psi_{,r}P_r \right), ~~~~
p_r  = - \sqrt{2} \psi_{,r}e^{\psi} P_r, ~~~~
p_{\phi} = 0 ~~~~ p_z = 0,  \\ 
&& E_{(r)} =  - \sqrt{2}\ e^{\psi} \psi_{,r}, ~~~~ E_{(z)} =  - \sqrt{2}\ e^{\psi} \psi_{,z} \label{eq_state}, 
\end{eqnarray}
with 
\begin{equation}
P_r = - \frac{2\sqrt{2}e^{\psi}}{r} \int  r \psi_{,z} e^{-\psi} \mathrm{d}r, 
\end{equation}
where  it was found that the electric potential $\chi$ can be written as $\chi =  \sqrt{2}\left( e^{\psi} - 1 \right)$ and  the metric function $\psi$   satisfies the following equation
\begin{equation}
\nabla^2 \psi - \nabla \psi \cdot \nabla \psi = 0,
\end{equation}
which can be cast as the Laplace equation $\nabla^2 \left( e^{-\psi} \right) = 0$. Furthermore, as we requiere the space-time to be asymptotically flat, the function $e^{-\psi}$ must be 1 at infinity, so we can write 
\begin{equation}
e^{-\psi} = 1 - U \, , \label{psi U}
\end{equation}
where $U$ is any solution of the Laplace equation that vanishes at infinity. Accordingly, equation (\ref{psi U}) allows us to write the physical quantities of the disk in terms of the function $U$, so we have
\begin{eqnarray}
\nonumber  \sigma &=&\frac{4}{(1-U)^2}\left[ U_{,z}-\frac{U_{,r}}{(1-U)r} \int \limits_{0}^{\infty} U_{,z} r \mathrm{d}r \right] \, , \\ \nonumber \\ 
 p_r &=& \frac{4U_{,r}}{r(1-U)^3}\int \limits_{0}^{\infty} U_{,z}r \mathrm{d}r \, , ~~~~~ 
 P_{(r)} =  \frac{ 2 \sqrt{2}}{(1 - U)^2 r} \int \limits_{0}^{\infty} U_{,z} r \mathrm{d}r \, ,  \\ \nonumber \\
\nonumber E_{(r)} &=& -   \frac{\sqrt{2} U_{,r} }{(1-U)^2} \, , ~~~~~
 E_{(z)} = -  \frac{\sqrt{2}U_{,z}}{(1-U)^2} \, . 
\end{eqnarray}
It can be seen that, the physical features of the disk depend entirely on the election of the solution of Laplace equation $U$. 



\section{A Particular Family of Polarized Disks}
\label{sec:PFPD}

Considering the axial symmetry of the system, we take as solution of Laplace equation the following function
\begin{equation}
U_{n} = -\sum^n_{l=0} \frac{C_l P_l\left( z/R \right)}{R^{l+1}}  \, ,  \label{U(PL)}
\end{equation}
with $C_l$ constants, $P_l(\cos \theta)$ the Legendre polynomials and $R=\sqrt{r^2+z^2}$. As we can see, this solution is zero once $R$ tends to infinity, which guarantees together with (\ref{psi U}) that the spacetime is asymptotically flat. Now, in order to introduce the discontinuities in the metric tensor  and in the electromagnetic four potential, we apply the  displace, cut and reflect method given by the the transformation 
\begin{equation}
z \rightarrow |z| + a \, , \label{metododesplazamiento}
\end{equation}
being $a$ a positive constant, to the function $U_n$. Accordingly, we now have
\begin{equation}
U_{n} = -\sum^n_{l=0} \frac{C_l P_l \left((|z| + a)/R \right)}{R^{l+1}}  \, ,  \label{U(PL)transf}
\end{equation}
with $R=\sqrt{r^2+(|z| + a)^2}$. Hence, the general expressions for the physical features of the disk are given by 
\begin{eqnarray}
 \nonumber \sigma &=&  \frac{4 \sum\limits_{l=0}^n \xi_1  }{\left(1 + \sum\limits_{l=0}^n  \xi_4 \right)^2}  
  - \frac{ 4\sum\limits_{m=0}^n  \xi_2   \int \limits_{0}^{\infty}  \sum\limits_{l=0}^n  \xi_3 r \mathrm{d} r }{\left(1+ \sum\limits_{l=0}^n \xi_4 \right)^3 } \, ,  \\ 
  p_r &=& \frac{ \sum\limits_{m=0}^n  \xi_2  \int \limits_{0}^{\infty}  \sum\limits_{l=0}^n \xi_3  r \mathrm{d}r  }{\left( 1+ \sum\limits_{l=0}^n \xi_4 \right)^3} ~~~~~ 
P_{(r)} = -  \frac{ 2 \sqrt{2} \int \limits_{0}^{\infty}   \sum\limits_{l=0}^n \xi_3 r \mathrm{d}r }{\left( 1 +  \sum\limits_{l=0}^n \xi_4 \right)^2 r}  \, ,  \\ 
\nonumber E_{(r)}&=& - \frac{\sqrt{2} r  \sum\limits_{l=0}^n  \xi_5  }{\left( 1 + \sum\limits_{l=0}^n \xi_4 \right)^2}, ~~~~~ E_{(z)}=  -  \frac{\sqrt{2} \sum\limits_{l=0}^n  \xi_1 }{\left( 1 + \sum\limits_{l=0}^n \xi_4 \right)^2}, 
\end{eqnarray}
with
\begin{eqnarray}
\nonumber &&\xi_1 = C_l(l+1) P_{l+1}(a/R_0)/R_0^{(l+2)}, ~~~\xi_2 =  C_{m}  P_{m+1}^{'}(a/R_0)/R_0^{(m+3)}  \, ,\label{xi_2}  \\ \nonumber \\
\nonumber &&\xi_3 = C_l(l+1) P_{l+1}(a/R_0) / R_0^{(l+2)}, ~~~ \xi_4 = C_l P_l(a/R_0)/R_0^{l+1}, ~~~ \xi_5 =  C_l  P_l^{'}(a/R_0)/R_0^{(l+3)} \, , \label{xi_5} 
\end{eqnarray}
and $R_0 = \sqrt{r^2 + a^2}$. In order to obtain a particular solution, it is necessary to specify the value $n$ and the constants $C_l$. 

As an example of electrically polarized thin disks, we consider the solution of the Laplace equation corresponding to the first term in the equation (\ref{U(PL)}), that is, when $n=0$
\begin{equation}
U_{0} = -\frac{C_0}{R}  \,, \label{Ucero}
\end{equation}
with $R=\sqrt{r^2+(|z| + a)^2}$. In consequence  the a-dimensional surface energy density  $\bar{\sigma}$, the radial pressure $\bar{p_r}$ and the radial component of the polarization vector $\bar{P}_{(r)}$ can be written as 
\begin{eqnarray}
  \bar{\sigma}  &=& \frac{\sigma}{\sigma_{0}} = \left(  \frac{1 + \bar{C_{0}}}{\bar{R_0} + \bar{C_{0}}} \right)^3, ~~~ 
 \bar{p_{r}} = \frac{p_r}{p_{r_0}} = \frac{1}{\bar{R_0}}\left( \frac{1 + \bar{C_{0}}}{\bar{R_0} + \bar{C_{0}} }\right)^3 ,~~~   \bar{P}_{(r)} =  \frac{- 2 \sqrt{2} \bar{C_{0}} \bar{R_0} }{(\bar{R_0} + \bar{C_{0}})^2 \bar{r}} \, ,
 \end{eqnarray} 
where the quantities $r = a \bar{r}$ and $C_{0} = a \bar{C_{0}}$, $R_0 = a \bar{R_0} $ were introduced in order to dimensionless the state variables of the system. Here $\sigma_{0} = \sigma|_{(r=0)}$, $p_{r_0} =  p_r|_{(r=0)}$.  Moreover, if we demand $\sigma$ and $p_r$ to satisfy the energy conditions \cite{RelativistikToolkit,Pimentel:2016jlm,Pimentel:2016hyo}, in order to obtain well behaved solutions, the value of the constant $ C_0$ must be $C_0=a$, which implies that $\bar{C_{0}} = 1$.

On the other hand, the electric potential and the components of the electric field are given by the following expressions 
\begin{equation}
\chi =  \frac{ -\sqrt{2} C_{0}  }{ R + C_{0}}, ~~~~~ E_{(r)} = - \frac{2 r C_0 }{ R( R + C_0)^2 }, ~~~~~ E_{(z)} = - \frac{2 z C_0 }{ R( R + C_0)^2 }.
\end{equation}
As we can see, if we even the electrical potential $\chi$ to a constant $\Gamma$ and solve for $\bar{r}$, we obtain an equation for the equipotential lines 
\begin{equation}
  \bar{r} = \pm \sqrt{ \nu  - (|\bar{z}| + 1)^2  }, ~~~~~ \nu = a^2 \bar{C_{0}}\left( \frac{\sqrt{2}}{\Gamma + 1} \right),   
\end{equation} 
where $z=a \bar{z}$. In the same way, in order to get the electric field lines, we replace the expressions for $E_{(r)}$ and $E_{(z)}$ into the following equation
$$
\frac{dz}{E_{(z)}} = \frac{dr}{E_{(r)}},
$$
and solve again for $\bar{r}$
\begin{equation}
  \bar{r} = \omega( |\bar{z}| + 1 ), ~~~~~ \omega = \sqrt{2\left( \frac{\bar{C_0}}{\bar{\kappa}} \right)^2-1}, 
\end{equation}  
where $\kappa = a\bar{\kappa}$.

\begin{figure}[h]
\centering
\includegraphics[width=0.32\textwidth ]{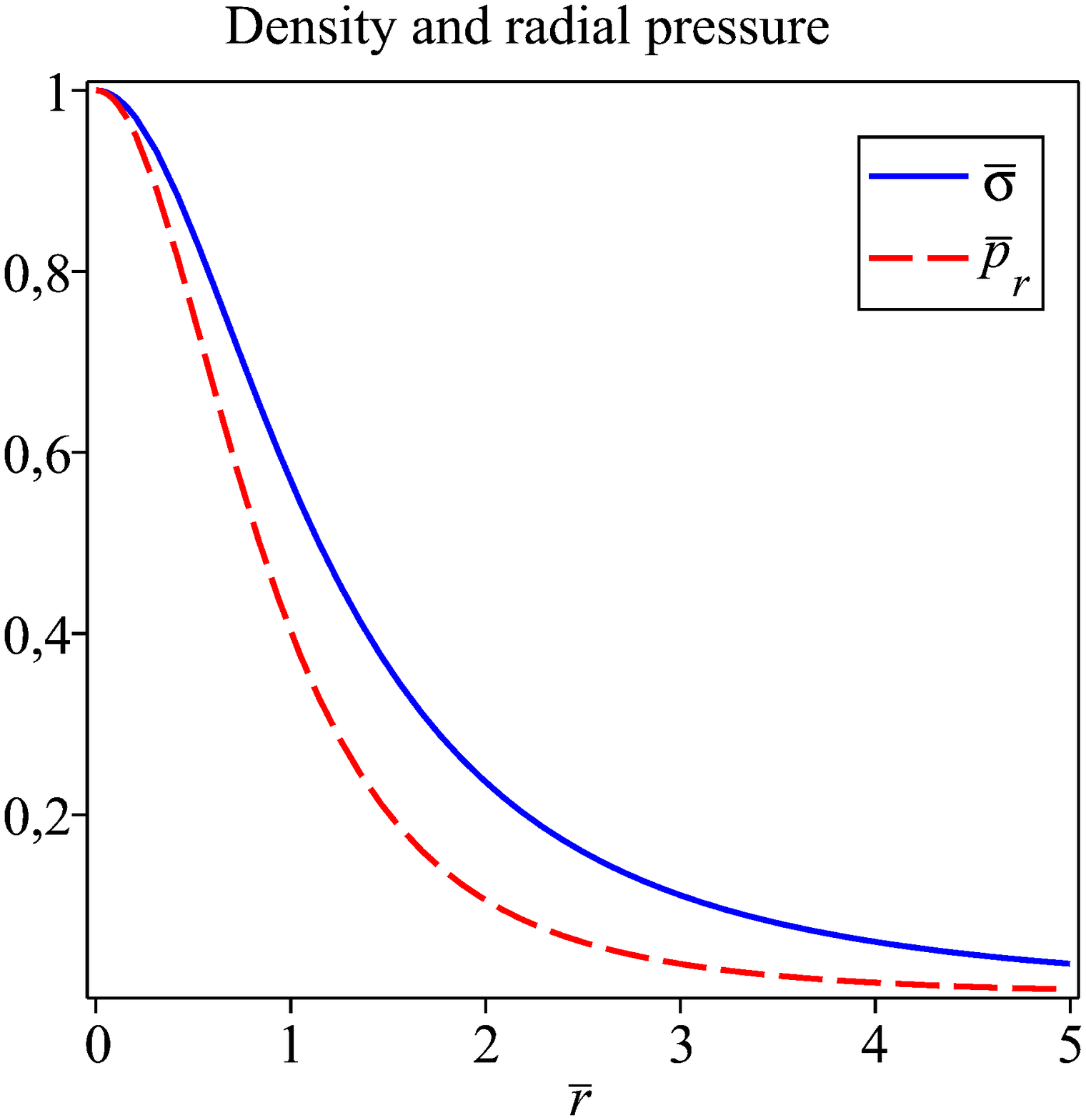}   
\includegraphics[width=0.32\textwidth]{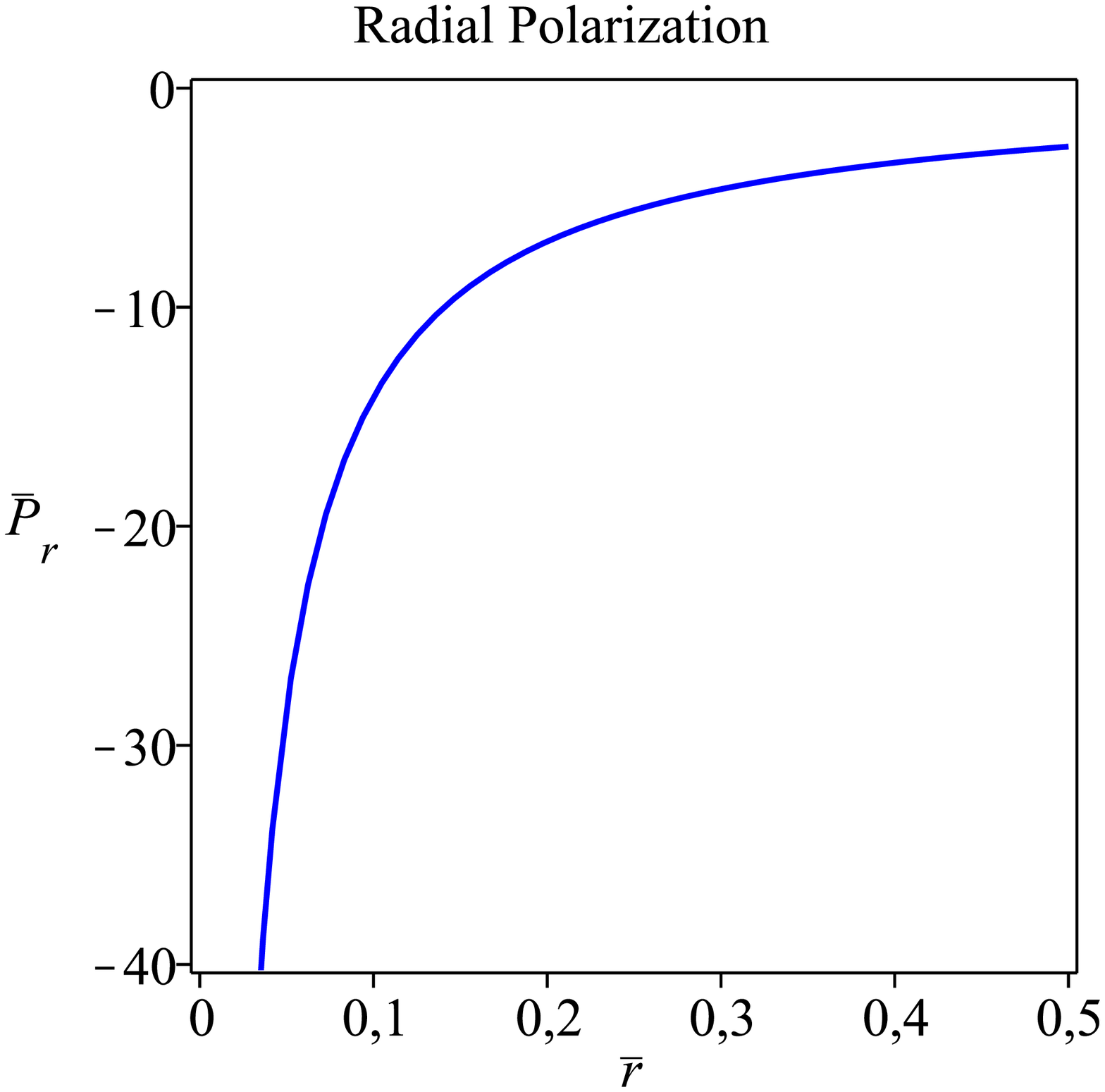} 
\includegraphics[width=0.32\textwidth]{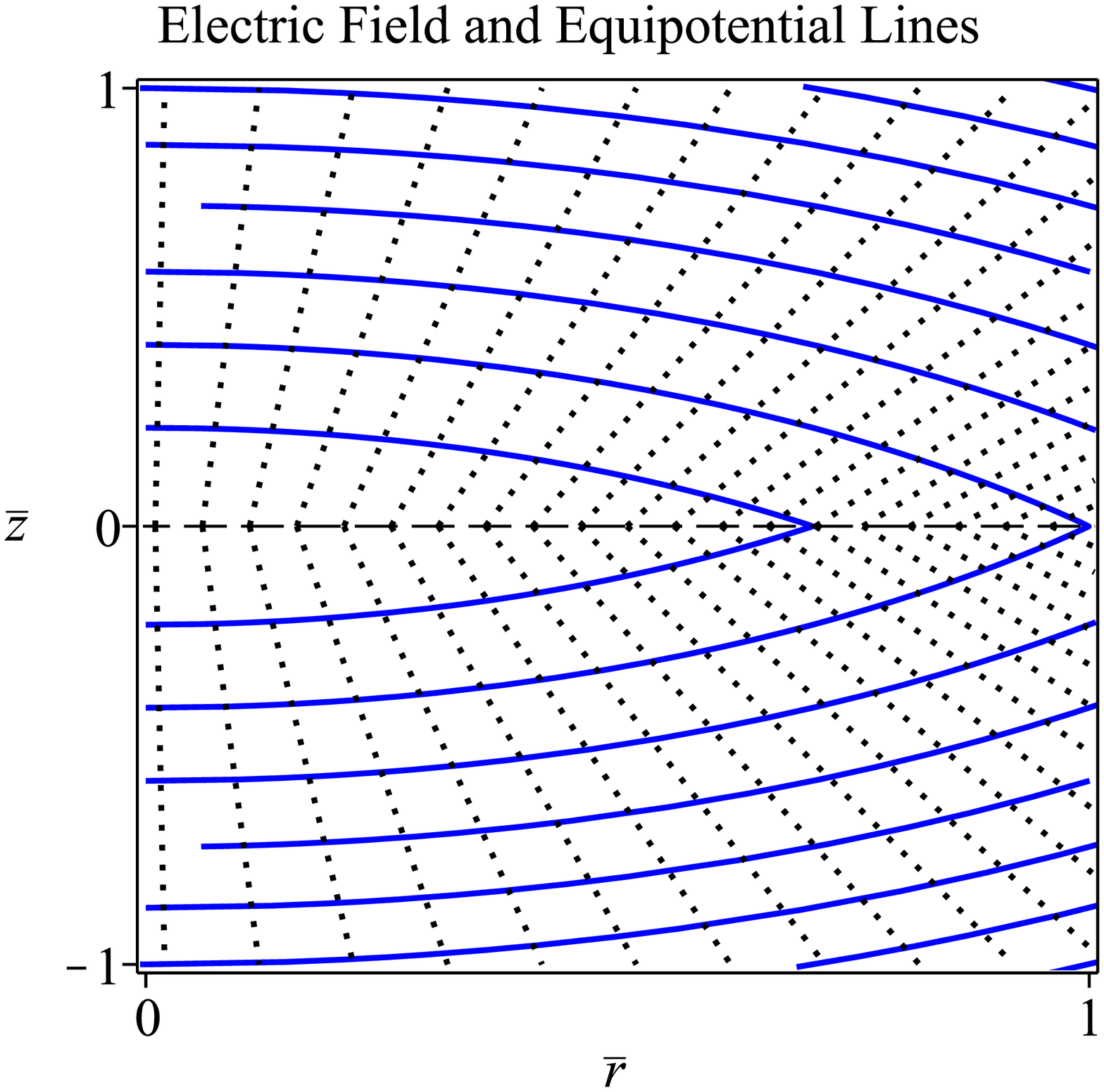} 
\caption{Physical features of the polarized disk $n=0$ with $\bar{C_0}=1$. } \label{Graficas_DP_n0}
\end{figure}

In figure \ref{Graficas_DP_n0}, we show the dimensionless profiles for the surface energy density $\bar{\sigma}$, radial pressure $\bar{p}_{r}$ (left panel),  radial polarization $\bar{P}_r$ (middle panel), the electric field and the equipotential lines (right panel) as a function of $\bar{r}$ for $\bar{C_0} = 1$. From these plots, it can be observed that $\bar{\sigma}$ and $\bar{p}_{r}$ are everywhere positives, the maximum values occur at the center of the disk and when $\bar{r}$ increases they decrease. On the other hand, the electric polarization vector $\bar{P_r}$ tends to zero when $\bar{r}$ tends to infinity and has a singularity at the center of the disk due to the form of function $U_0$, which is a solution of the Laplace equation. 

\section{ROTATION CURVES}
\label{sec:RC}

From the astrophysical point of view, an important aspect that can be studied in this system is the circular geodesics along the plane of the disk ($z=0$),  in order to qualitatively determine its resemblance with the observed rotation curves. For the line element (\ref{ds2}) and assuming $r$ constant, such movement has a velocity given by
\begin{equation}
u^\alpha = u^t \left( 1 , 0,\Omega , 0 \right) \, ,
\end{equation}
where $\Omega = u^\phi/u^t $ is the angular velocity. By imposing the normalization condition $g_{\alpha \beta} u^\alpha u^\beta = -1 $, we obtain the relation
\begin{equation}
\left( u^t \right)^2 = \frac{e^{-2\psi}}{1 - {V_c}^2 } \, ,
\end{equation}
where ${V_c} = e^{-2\psi}r\Omega$ is the circular o tangential velocity. On the other hand, from the geodesic equation
\begin{equation}
\frac{d u_\alpha}{d\tau} = \frac{1}{2} g_{\mu \nu , \alpha} u^\mu u^\nu \, ,
\end{equation}
the angular velocity can be written as 
\begin{equation}
\Omega = \frac{\psi_{,r} \ e^{4\psi}}{r ( 1 - r \psi_{,r}  ) } \, , 
\end{equation}
and then we get the following expression for the circular velocity in terms of the function $U$
\begin{equation}
{V_n}_{c} = \sqrt{\frac{ r {U_n}_{,r} }{ 1 - U_n - r {U_n}_{,r} }   }.
\end{equation}

\begin{figure*}[h]
\centering
$$\begin{array}{cc}
\includegraphics[width = 0.5\textwidth]{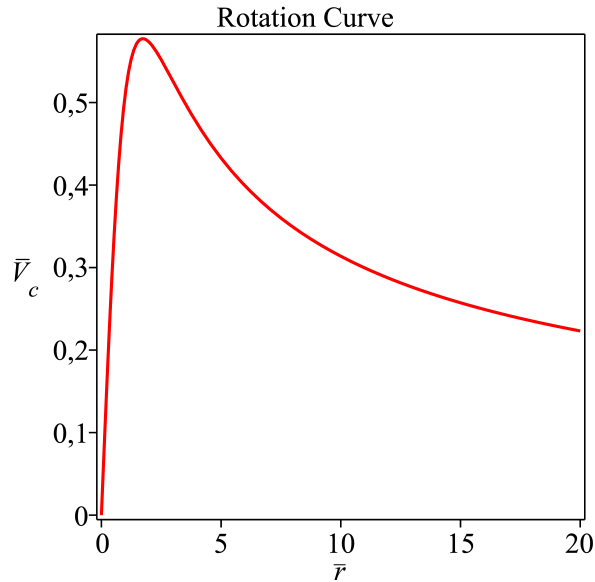}&
\end{array}$$
\caption{Rotation curves for $n = 0$} \label{Graficas_curvasrotacion}
\end{figure*}

For the cases above mentioned, that is $n=0$, the rotation curve is given as
\begin{eqnarray}
{V_0}_c = \bar{r} \sqrt{\frac{ \bar{C_0} }{ \bar{R_0}^3 + \bar{C_0} } },
\end{eqnarray} 
The rotation curve is displayed in figure \ref{Graficas_curvasrotacion},  and its behavior is similar with the predicted rotation curves from Newtonian gravity for spiral galaxies \cite{Andromeda_Galaxy}. 

\section{Conclusions}

A infinite family of exact solutions of the Einstein-Maxwell equations was obtained for an axially symmetric conformastatic spacetime. The solution describes  thin disks electrically polarized. This model was obtained by using the distributional approach of tensors, and introducing finite discontinuities in the first normal derivatives of the metric tensor and the electromagnetic four potential. We analyzed the first model of the  family, that is, the model $n=0$, finding that the disk behaves in a reasonable way, that is, the functions are everywhere positive, smooth, decay when the distance from the center increases and show no singularities except for the electrically polarization vector. Despite of the infinite extension of the disk the physical features are concentrated in the central region as we could expect from a physical system. This work serves as a starting point for the study of electromagnetic polarized matter within the exact solutions of disk-like configurations. 

\section{Acknowledgment}
A. N wants to thanks the financial support from COLCIENCIAS and Universidad Industrial de Santander. F.D.L-C gratefully acknowledges the financial support from Vicerrector\'ia de Investigaci\'on y Extensi\'on,  Universidad Industrial de Santander, grant number 1822. G. A. G. was supported in part by VIE-UIS, under Grants No. 1347 and No. 1838, and by COLCIENCIAS, Colombia, under Grant No. 8840.

\section*{References}

\providecommand{\newblock}{}

\end{document}